
\magnification=1200
\baselineskip=1.8\normalbaselineskip
\nopagenumbers
\centerline{\bf Limiting Behavior of Solutions to the}

\centerline{\bf Einstein-Yang/Mills Equations}

\centerline{\bf by J. A. Smoller and A. G. Wasserman}

Spherically symmetric solutions to the Einstein-Yang/Mills equations with gauge
group $SU(2)$ have been investigated by many authors, c.f. [1 - 8].  These
equations take the form of a second order ordinary differential equation for
the
connection coeffici
ent $w ,$ coupled to a first order equation for the metric coefficient $A ,$
(together with a de-coupled first-order equation for the metric coefficient
$T$). Particle-like solutions; ie, globally defined regular solutions, were
first found by Bartnik and
 McKinnon, [1] and and were rigorously shown to exist in [4].  The existence of
infinitely-many such solutions, one in each nodal class proved in [5]; see also
the paper [3] by K\"unzle and Masood-ul-Alam.  Black hole solutions; i.e.,
solutions defined fo
r all $r \ge \rho > 0 ,$ with $A(\rho) = 0 ,$ were investigated in [3], as well
as in the paper [2], by Bizon.  It was proved in [6], that for any $\rho > 0 ,$
there exist an infinite number of black hole solutions, one in each nodal
class.

In the paper [7], the authors considered the limiting behavior of a sequence of
particle-like solutions, indexed by nodal class, $n ,$ as $n \to \infty .$ The
results of that paper are hereby corrected and improved to include the limiting
behavior of blac
k hole solutions also.  The following theorem cover both cases:

\noindent \underbar{\bf Theorem}: If $\displaystyle{\{\Lambda_n (r) \equiv (w_n
(r) , w^\prime_n (r) , A_n (r) , r)\} ,}$ is a sequence of radially symmetric
black-hole solutions to the Einstein-Yang/Mills equations with event horizon
$\rho$ (particle-lik
e solutions if $\rho = 0$), indexed by nodal class, $n ,$ and if $n$ tends to
$\infty ,$ then for
$r > 1 ,$
\hfil
\break
$\displaystyle{\lim_{n \to \infty} \Lambda_n (r) = (0 , 0 , A(r) , r) ,}$ where
$\displaystyle{A (r) = 1 - {{2} \over {r}} + {{1} \over {r^2}}}$ if $\rho \le 1
,$ and
\hfil
\break
$\displaystyle{A(r) = 1 - {{(\rho + \rho^{-1})} \over {r}} + {{1} \over
{r^2}}}$
if $\rho \ge 1 .$ Moreover the convergence is uniform on bounded $r$-intervals
in $(\rho , \infty) .$
\vfil
\eject
One interesting corollary of this theorem is that for any (event horizon) $\rho
\ge 0 ,$ and any sequence of solutions for which  $n ,$ the nodal class tends
to
$\infty ,$ the (ADM) masses must converge to 2, if $\rho \le 1 ,$ and to $\rho
+
\rho^{-1} \ ,
 \ {\hbox{if}} \ \rho > 1 .$

The results in sections 2, 3, and 5 in [7] are correct as stated, as are Lemmas
4.2 and 4.5.  The other results in \S4 are incorrect, and \S6 is no longer
relevant, in view of the above theorem.

\centerline{\bf REFERENCES}

\item{1.}{Bartnik, R., and Mckinnon, Phys. Rev. Lett. 61, (1988), 141-144.}

\item{2.}{Bizon, P., Phys. Rev. Lett., 64, (1990), 2844-2847.}

\item{3.}{K\"unzle, H. P., and Masood-ul-Alam, A.K.M., J. Math. Phys., 31,
(1990), 928-935.}

\item{4.}{Smoller, J., Wasserman, A., Yau, S.-T., and McLeod, B., Comm. Math.
Phys., 143, (1991), 115-147.}

\item{5.}{Smoller, J., and Wasserman, A., Comm. Math. Phys. 151, (1993),
303-325.}

\item{6.}{Smoller, J., Wasserman, A., and Yau, S.-T., Comm. Math. Phys., 154,
(1993),
\hfil
\break
377-401.}

\item{7.}{Smoller, J., and Wasserman, A., Comm. Math. Phys., 161, (1994),
365-389.}

\item{8.}{M. S. Volkov and D. V. Galtsov, Sov. J. Nucl.Phys. 51, (1990), 1171.}

\hskip 3in Mathematics Department

\hskip 3in University of Michigan

\hskip 3in Ann Arbor, MI \ 48109-1003

\hskip 3in U.S.A.

\end